\documentclass[twocolumn,showpacs,preprintnumbers,amsmath,amssymb,prl,epsf]{revtex4}
\usepackage{graphicx}
\begin{document}
\title{Generation of different Bell states\\
within the SPDC phase-matching bandwidth}
\author{G. Brida, M. Genovese and L.~A.~Krivitsky}
\affiliation{Istituto Nazionale di Ricerca Metrologica, Strada delle
Cacce 91, 10135 Torino, Italy}
\author{M.~V.~Chekhova }
\affiliation{Department of Physics, M.V.Lomonosov Moscow State
University,\\
Leninskie Gory, 119992 Moscow, Russia} \vskip 24pt

\begin{abstract}
\begin{center}\parbox{14.5cm}
{We study the frequency-angular lineshape for a phase-matched
nonlinear process producing entangled states and
 show that there is a continuous variety of maximally-entangled states
 generated for different mismatch values within the natural
bandwidth. Detailed considerations are made for two specific methods
of polarization entanglement preparation, based on type-II
spontaneous parametric down-conversion (SPDC) and on SPDC in two
subsequent type-I crystals producing orthogonally polarized photon
pairs. It turns out that different Bell states are produced at the
center of the SPDC line and on its slopes, corresponding to about
half-maximum intensity level. These Bell states can be filtered out
by either frequency selection or angular selection, or both. Our
theoretical calculations are confirmed by a series of experiments,
performed for the two above-mentioned schemes of producing
polarization-entangled photon pairs and with two kinds of
measurements: frequency-selective and angular-selective.}
\end{center}
\end{abstract}
 \maketitle \narrowtext
\vspace{-10mm}

\section{I. Introduction}

Spontaneous Parametric Down-Conversion (SPDC) provides the easiest
way of generating polarization-entangled two-photon states, which
are a fundamental resource for quantum communication or quantum
computation~\cite{NC}, studies on foundations of quantum
mechanics~\cite{prep}, quantum metrology~\cite{las}, quantum
imaging~\cite{qi}, etc.

Various experimental schemes can be used for this
purpose~\cite{prep}. The first efficient scheme suggested was the
one based on non-collinear type-II SPDC~\cite{Kwiat_rings}. In this
scheme, entangled two-photon states were generated in two angular
(wavevector) modes and two polarization modes
(polarization-wavevector entanglement). Later, 'interferometric'
schemes were also proposed, where two crystals were involved or two
passages of the pump beam through the same crystal. In most of these
schemes, the set of modes for entanglement also included two
polarization modes and two wavevector modes. Among numerous
configurations, one should mention  schemes with type-II crystals
specially designed for the operation in the femtosecond-pulsed
regime~\cite{Branning,Kim} and the scheme with two type-I
crystals~\cite{Kwiat}.  Less common but very useful for certain
applications are interferometric schemes that provide entanglement
in a set of two frequency modes and two polarization modes
(polarization-frequency entanglement)~\cite{Sergei+Yoon-Ho,Yura}.

We would like to emphasize that in all such experiments the goal was
to achieve a certain Bell state generated within the whole bandwidth
given by phase matching. By changing the settings of the optical
elements, other Bell states can be obtained, but still for the whole
phase-matched range of wavelengths and wavevectors. In type-II-based
schemes, this is provided by inserting a birefringent optical
element compensating for the delay between orthogonally polarized
photons of an entangled pair ('e-o delay compensation')~\cite{Shih}.
Otherwise, entanglement is obtained only for the narrow frequency
domain around exact degeneracy, selected by means of narrow-band
filters. In type-I schemes, e-o delay compensation or filtering are
seemingly not necessary but, as we will show further, still
preferable since entanglement is not uniform over the whole
bandwidth and changes on the slopes of the SPDC phase-matching line.

In this paper, we are interested in the fine structure of the
phase-matching line from the viewpoint of various Bell states
generation. We find that without the e-o delay compensation, an
interesting effect occurs within the bandwidth for every
phase-matched nonlinear process: the same phase mismatch that is
responsible for the intensity decrease induces a phase shift between
the components of the  produced entangled state. In particular, we
will discuss in detail the SPDC case. In this case, we show that
various maximally-entangled states and, in particular, two different
Bell states are generated within the spectrum allowed by
phase-matching and that one can choose them by means of frequency or
angular selection. This occurs for all schemes of
polarization-entangled states generation - based on type-II phase
matching and based on the 'two type-I crystals' configuration - and
involves both frequency or angular lineshapes of SPDC. For example,
both for type II and `two-type I' near-collinear production one
finds inside the natural bandwidth maximally entangled states (${|HV
\rangle + e^{i \phi} | VH \rangle \over \sqrt 2} $ and ${|HH \rangle
+ e^{i \phi} | VV \rangle \over \sqrt 2} $ respectively), whose
relative phase between components, $\phi$, ranges from 0 to $\pi$.

After a detailed theoretical discussion, we present experimental
data demonstrating it in various cases.

 The paper is organized as follows. In Section II, we
present the theory of the effect and describe a general scheme for
its observation. In particular, the two-dimensional
(frequency-angular) spectrum of SPDC is presented, with the
calculated domains of various Bell states generation shown. Sections
III and IV are devoted to the experimental observation of the
effect. Section III is focused on the frequency spectrum of
collinear SPDC, both for type-II and 'two type-I' schemes. The
effect is observed in two ways: by scanning the frequency of one of
the entangled photons and by scanning the arrival-time difference of
the pair, after transmitting it through an optical fibre. Section IV
contains experimental results on the SPDC angular spectrum. In this
group of measurements, also carried out both for type-II phase
matching and the 'two type-I configuration', switching between
different Bell states is performed by scanning the wavevector
direction of one of the two photons. Finally, in Section V we draw
the conclusions and discuss possible applications of the observed
effect.

\section{II. Theory}

Let us consider polarization-entangled two-photon state generated
via a phase-matched process, which can be, for instance, SPDC or
four-wave mixing~\cite{FWM}. Ignoring the vacuum component, we can
write  the state in the general form (see, e.g.,~\cite{DNK}),

\begin{equation}
\left|\Psi\right>=\int d\bold k d\bold {k'}\int d\bold r
F_{NL}(\bold r)e^{i\bold {\Delta}(\bold{k,k'})\bold r}a_{P,\bold
k}^{\dagger}a_{P',\bold {k'}}^{\dagger}\left|\hbox{vac}\right>,
\label{Psi}
\end{equation}
(with polarizations $P=H,P'=V$ for type II and $P=P'$ for type I
phase matching, respectively), where $\bold{\Delta}(\bold k,
\bold{k'})$ is the phase mismatch depending on wavevector and
frequency modes $\bold k, \bold{k'}$, in which a pair is created,
$a_{P\bold {k,k'}}^{\dagger}$ are photon creation operators and
$F_{NL}(\bold r)$ is the factor containing nonlinearity and the pump
(pumps) field distribution over the nonlinear medium. Integration
over the volume turns $F_{NL}(\bold r)$ into its Fourier transform,
$F_{NL}(\bold {k, k'})$, but there remains a phase factor, depending
on $\bold k, \bold{k'}$. If this phase factor is not symmetric
w.r.t. the exchange $\bold k \leftrightarrow\bold{k'}$, as it indeed
happens in nearly all cases, then the phase between the two
components of an entangled state generated at the output will change
within the phase-matched bandwidth.

Let us now consider the case where polarization-entangled states are
generated through SPDC. We will separately focus on the situations
of type-II phase matching (in a single crystal) and of type-I phase
matching (in two crystals pumped by the same pump and the optic axes
oriented in orthogonal planes).

\subsection{II.1 Type II phase matching}

Consider first polarization-entangled photons produced via type-II
SPDC in a homogeneous crystal of length $L$ from a cw pump with
relatively large beam diameter $a>>L\tan\theta$, $\theta$ being the
typical angle of scattering, so that exact transverse phase matching
can be assumed~\cite{BigDNK}. Integration over the volume leads to a
delta-function of the transverse mismatch,
$\delta^{(2)}(\bold{k_{\bot}}+\bold{k'_{\bot}})$, and due to the
strict frequency correlation between the two photons, integration
over the modes $\bold{k'}$ in Eq.(\ref{Psi}) disappears. In the
two-photon amplitude, only the longitudinal part of the wavevector
mismatch is relevant, $\Delta_z(\bold k)$, where $z$ is the
direction along the pump wavevector. In terms of experimentally
measurable quantities, integration in $\bold{k}$ means integration
w.r.t. the angle of scattering and frequency of one of the photons
belonging to the entangled pair. Further, assuming that the signal
and idler photons have wavevectors belonging to the plain of the
optic axis and labeling them as $o$ and $e$ (ordinary and
extraordinary), we can write the state as

\begin{eqnarray}
\left|\Psi\right>=\iint d\theta\ d\Omega
e^{i\frac{\Delta_zL}{2}}\hbox{sinc} \left(\frac{\Delta_z L}{2}
\right)\nonumber
\\
\times
a_e^{\dagger}(\theta,\omega_0+\Omega)a_o^{\dagger}(-\theta,\omega_0-\Omega)\left|\hbox{vac}\right>,
\label{Psi_II}
\end{eqnarray}
where  $\omega_0$ is the central frequency of SPDC spectrum, the
$\hbox{sinc}$ function is defined as $\hbox{sinc}(x)\equiv\frac{\sin
x}{x}$, and the longitudinal phase mismatch has the form

\begin{eqnarray}
\Delta_z\equiv\Delta_z(\theta,\Omega)=k_{e}(\theta,\omega_0+\Omega)\cos\theta\nonumber
\\+ k_o(-\theta,\omega_0-\Omega)\cos\theta-k_p, \label{mismatch}
\end{eqnarray}
$k_p, k_{e}$, and $k_o$ being the wavevectors of the pump, signal
(extraordinary), and idler (ordinary) photons, respectively. From
this, one can see that with the change in the mismatch $\Delta_z$,
determining the lineshape of SPDC, there occurs a simultaneous
change in the phase factor standing by the state. If we take into
account both positive and negative values of the mismatch, we find
that different types of two-photon states can be generated within
the spectrum allowed by phase-matching. For instance, the state
generated at exact phase matching ($\Delta_z=0$) is a factorable
one, but it can be turned into the Bell state $\left|\Psi^+\right>$
by means of a beamsplitter. At mismatch values $\Delta_z=\pm\pi/L$,
which can be realized by means of frequency offset from degeneracy,
or angular offset, or both, the state already becomes another Bell
one, the singlet state $\left|\Psi^-\right>$. At intermediate
mismatch values, various maximally entangled states
 ${|HV \rangle + e^{i \phi} | VH \rangle \over \sqrt
2}$ are generated, their phases $\phi$ being between $0$ and $\pi$.

Expanding the mismatch (\ref{mismatch}) around the collinear
frequency-degenerate point, where exact phase matching is assumed,
up to linear terms in frequency and angular offsets, we obtain

\begin{equation}
\Delta_z(\theta,\Omega)\simeq D\Omega+B\theta, \label{expand}
\end{equation}
where $D\equiv\frac{dk_{e}}{d\Omega}-\frac{dk_o}{d\Omega},
B\equiv\frac{dk_{e}}{d\theta}$.

One can observe that the singlet state $\left|\Psi^-\right>$ is
generated for a continuous set of frequencies and angles
$\theta,\Omega$, for which $\Delta_z(\theta,\Omega)=\pm\pi/L$. This
is shown in Fig.1a where the frequency-angular spectrum of SPDC is
calculated for the case of collinear frequency-degenerate type-II
phase matching in a $0.5$ mm BBO crystal. The solid line corresponds
to $\Delta_z(\theta,\Omega)=0$, i.e. $\left|\Psi^+\right>$
production; dashed lines correspond to
$\Delta_z(\theta,\Omega)=\pm\pi/L$, i.e. $\left|\Psi^-\right>$
generation. Note that at angles and frequencies corresponding to
$\left|\Psi^-\right>$ generation, the intensity of SPDC is only
about $0.4$ times as large as the intensity at the maximum.

\noindent\textbf{Frequency spectra.} For the sake of simplicity let
us start by considering the collinear case, i.e. the angle between
the photons $\theta=0$. In experiment, this can be provided by a
pinhole inserted into the SPDC beam. In this case the two-photon
state (\ref{Psi_II}) can be written as a frequency expansion:

\begin{eqnarray}
|\Psi\rangle=\int_0^{\infty}{d\Omega}\hbox{sinc}(\tau_0\Omega)[a^{\dagger}_{H}(\omega_0+\Omega)
a^{\dagger}_{V}(\omega_0-\Omega)e^{i\tau_0\Omega}
\nonumber\\
+a^{\dagger}_{V}(\omega_0+\Omega)
a^{\dagger}_{H}(\omega_0-\Omega)e^{-i\tau_0\Omega}]|\hbox{vac}\rangle,
\label{IIfreq}
\end{eqnarray}
where $\tau_0\equiv{DL/2}$ is the e-o delay for a pair born at the
center of the crystal. Here, we assume the e-photon to be
horizontally polarized; the integration runs only on positive
frequency offsets.

We see that while at the degenerate frequency, the state
$\left|\Psi^+\right>$ is created (if the photon flux is split in two
beams by a beamsplitter), the state generated at
$\Omega=\pm\pi/2\tau_0$, i.e., at approximately half-maximum of the
intensity, is the singlet state $\left|\Psi^-\right>$.

If a birefringent plate is inserted into the beam after the crystal,
then an additional e-o delay $\tau$ will be added to $\tau_0$ in
Eq.~(\ref{Psi_II}). The case $\tau=-\tau_0$ is well known: it means
that the e-o delay arising in the crystal is completely compensated.
Then, for the whole bandwidth allowed by phase matching, one
produces a $\left|\Psi^+\right>$. More interesting for our
considerations is the case where $\tau$ has the same sign as
$\tau_0$. In this case the change of the relative phase between the
two state components in Eq.(\ref{Psi_II}) becomes more rapid than
the intensity variation determined by the phase matching, and both
$\left|\Psi^+\right>$ and $\left|\Psi^-\right>$ are generated in
several domains within the linewidth.

\noindent\textbf{Angular spectra.} Similarly, if one selects the
frequencies of the two photons, for instance, corresponding to exact
degeneracy, then the state can be written as the angular spectrum,

\begin{eqnarray}
|\Psi\rangle=\int_0^{\infty}{d\theta}\hbox{sinc}
\left(\frac{B\theta}{2} \right)[a^{\dagger}_{H}(\theta)
a^{\dagger}_{V}(-\theta)e^{i\frac{B\theta}{2}}
\nonumber\\
+a^{\dagger}_{V}(\theta)
a^{\dagger}_{H}(-\theta)e^{-i\frac{B\theta}{2}}]|\hbox{vac}\rangle,
\label{IIang}
\end{eqnarray}

Hence, the same 'fine structure' of Bell states is present within
the angular lineshape of SPDC. Similarly to the frequency case, this
structure can be removed or made finer  by, respectively, reducing
(compensating) or increasing the anisotropy. Compensation of the
anisotropy can be achieved if an anisotropic plate is placed after
the crystal; for example, a similar crystal with the length twice
smaller and the optic axis rotated $180^o$ with respect to the pump
axis. However, in this case one should avoid SPDC generation in the
second crystal. This can be easily achieved by either eliminating
the pump after the first crystal or slightly detuning the
orientation of the optic axis of the second crystal from exact phase
matching. In the last case the resulting change in the $B$ value can
be compensated for by choosing a crystal of length slightly
differing from $L/2$.

It is worth mentioning that the fine structure is absent if the
angle is scanned in the plane orthogonal to the plane of the optic
axis.

\subsection{II.2 Type I phase matching}

Consider now generation of polarization-entangled photon pairs by
placing two type-I crystals one after the other into a common pump
beam, the optic axes of the crystals being in orthogonal planes. In
the following we consider the case where the pump  gives equal
contributions to the SPDC intensities in both crystals (i.e. its
polarization is at $45^o$ w.r.t. the optic axis planes of the
crystals). For this configuration, there are two substantial
differences from the type-II case considered above. First, the
wavevector mismatch should be expanded up to quadratic terms in the
angle and frequency offsets, since the linear terms cancel each
other:

\begin{equation}
\Delta_z(\theta,\Omega)\simeq
\frac{d^2k_{o}}{d\Omega^2}\Omega^2-k_o\theta^2. \label{expandI}
\end{equation}

The form of the mismatch is the same for both crystals. Second, an
ordinarily polarized pair born in the first crystal acquires a
frequency- and angle-dependent phase factor in the second crystal,
where it is extraordinarily polarized, and the state generated at
the output is

\begin{eqnarray}
\left|\Psi\right>=\iint_0^{\infty} d\theta\ d\Omega \hbox{sinc}
 \left(\frac{\Delta_z L}{2} \right)\nonumber
\\
\times
[a_H^{\dagger}(\theta,\omega_0+\Omega)a_H^{\dagger}(-\theta,\omega_0-\Omega)\nonumber
\\+e^{i\phi(\theta,\Omega)}a_V^{\dagger}(\theta,\omega_0+\Omega)a_V^{\dagger}(-\theta,\omega_0-\Omega)]\left|\hbox{vac}\right>,
\label{Psi_I}
\end{eqnarray}
where, up to second-order terms in frequency and angular offsets,
$\phi(\theta,\Omega)=(\frac{d^2k_{e}}{d\Omega^2}\Omega^2-k_{e}\theta^2+\frac{d^2k_{e}}{d\theta^2}\theta^2)L$,
and the same assumptions on the pump are made as when deriving
Eq.(\ref{Psi_II}). The third term in the expression for
$\phi(\theta,\Omega)$ is much smaller than the other two and hence
can be omitted. As a result, up to the exchange $k_o\rightarrow
k_{e}$, the phase $\phi(\theta,\Omega)$ is equal to the 'mismatch
phase' $\Delta_zL$, and the same effect of 'switching' between
different maximally entangled states as for the type-II case takes
place for this configuration. However, while for the type-II phase
matching 'switching' occurred between the states
$\left|\Psi^+\right>$ and $\left|\Psi^-\right>$, this time the Bell
states involved are $\left|\Phi^+\right>$ and $\left|\Phi^-\right>$.
The structure of the frequency-angular spectrum in this case is
shown in Fig.1b, with the state $\left|\Phi^+\right>$ shown by solid
lines and the state $\left|\Phi^-\right>$ by dashed lines. It is
important to emphasize that, a fact that, to our knowledge has  not
been mentioned yet, also the `two-type I' scheme of Bell-states
preparation requires some filtering:  the Bell state produced on the
slopes of the spectrum is different from the state produced in the
middle.

From Fig.1b one can also observe that, when polarization-entangled
states are produced using 'two type-I' phase matching, transition
between various maximally entangled states occurs at angles and
frequencies where the phase mismatch changes rapidly. This domain is
rather narrow; in the most part of the frequency-angular spectrum,
only the state $\left|\Phi^+\right>$ is produced.

Unlike  type-II phase matching, here the 'fine structure' of the
angular lineshape appears regardless of the plane where the angular
scanning is made.

\noindent\textbf{Frequency and angular spectra.} The frequency
spectrum of the state in the collinear case is

\begin{eqnarray}
\left|\Psi\right>=\int_0^{\infty} d\Omega \hbox{sinc}\left
(\frac{d^2k_{o}}{d\Omega^2}\frac{\Omega^2L}{2} \right)\nonumber
\\
\times
[a_H^{\dagger}(\omega_0+\Omega)a_H^{\dagger}(\omega_0-\Omega)\nonumber
\\+e^{i\frac{d^2k_{e}}{d\Omega^2}\Omega^2L}a_V^{\dagger}(\omega_0+\Omega)
a_V^{\dagger}(\omega_0-\Omega)]\left|\hbox{vac}\right>,
\label{Psi_Ifreq}
\end{eqnarray}
and the angular spectrum in the degenerate case is

\begin{eqnarray}
\left|\Psi\right>=\int_0^{\infty} d\theta \hbox{sinc}
\left(\frac{k_{o}\theta^2L}{2} \right)\nonumber
\\
\times [a_H^{\dagger}(\theta)a_H^{\dagger}(-\theta)\nonumber
\\+e^{i k_{e}\theta^2 L}a_V^{\dagger}(\theta)
a_V^{\dagger}(-\theta)]\left|\hbox{vac}\right>. \label{Psi_Iang}
\end{eqnarray}

\section{III. Experiment: polarization-frequency entanglement}

The general experimental setup for exploring the frequency-angular
Bell states structure inside the bandwith
 is sketched in Fig.2. Two-photon
light generated in a 0.5 mm type-II crystal or a combination of 1 mm
type-I crystals was split using a non-polarizing beamsplitter and a
Glan prism was put into each output port. In one of the output
ports, angular or frequency selection was performed, the detector in
the other arm collecting all possible conjugate modes. Coincidences
of the two detectors were measured depending on the positions of the
Glan prisms and the frequency and angle selected.

In particular, in the specific experimental setup for testing
polarization-frequency Bell states, produced within the bandwidth of
collinear frequency-degenerate SPDC,  a diffractive-grating
monochromator with the resolution 0.8 nm was used as the
frequency-selective element. Since the pump was continuous-wave,
frequency selection for one of the photons of a pair automatically
selected the frequency of the correlated photon. To reduce the
contribution of accidental coincidences, a broadband interference
filter centered around 702 nm was placed after the beamsplitter.
Biphoton pairs were registered by two photodetection apparatuses,
consisting of red-glass filters, pinholes, focusing lenses and
avalanche photodiodes. The photocount pulses of the two detectors,
after passing through delay lines, were sent to the START and STOP
inputs of a time-to-amplitude converter (TAC). The output of the TAC
was finally addressed to a multichannel analyzer (MCA), and the
number of photocount coincidences of the two detectors was observed
at the MCA output.

In a modified version of this setup, an optical fibre was used as a
frequency-selective element. Indeed, as it was shown
in~\cite{Brida}, due to the group-velocity dispersion, an optical
fibre performs a Fourier transformation of the two-photon spectral
amplitude, which can be measured through the time distribution of
the signal-idler delay after the fibre. Thus,  in some of our
measurements the monochromator in one output port was replaced by an
optical fibre of length $1$ km inserted into the beam before the
beamsplitter. This specific set-up is interesting for application to
long-distance fiber quantum communication as the transmission fibre
takes the role of transforming the frequency selection to (very
easily implementable) time window selection.

\subsection{III.1 Type-II configuration.}

According to Eq.(\ref{IIfreq}), the frequency spectrum of the
type-II configuration contains a continuum of polarization-frequency
entangled states from $\left|\Psi^+\right>$ to
$\left|\Psi^-\right>$. A usual test of such states consists of
fixing one of the polarization prisms at an angle
$\Theta_1=45^{\circ}$ and measuring the coincidence counting rate as
a function of the other prism orientation $\Theta_2$ (polarization
interference). For Bell states $\left|\Psi^{\pm}\right>$, high
interference visibility (100\% in the ideal case) should be
observed, the coincidence counting rate depending on
$\Theta_1,\Theta_2$ as $R_c\sim\sin^2(\Theta_2\pm\Theta_1)$ and its
minimum  corresponding to $\Theta_2=\mp45^{\circ}$.

To demonstrate the generation of various polarization-entangled
states within the collinear type-II SPDC bandwidth, we measured the
coincidence counting rate versus the wavelength selected by the
monochromator, for two settings of the Glan prisms:
$\Theta_1=45^{\circ},\Theta_2=-45^{\circ}$ and
$\Theta_1=\Theta_2=45^{\circ}$ (Fig.3). The results fit well the
dependence, calculated in~\cite{Brida} as

\begin{eqnarray}
R_c(\Omega)=\hbox{sinc}^2(\Omega\tau_0)[\sin^2(\Theta_1+\Theta_2)\cos^2(\Omega\tau_0)
\nonumber
\\
+\sin^2(\Theta_1-\Theta_2)\sin^2(\Omega\tau_0)]. \label{thetas}
\end{eqnarray}

As it follows from Eq.(\ref{thetas}) and is confirmed by
experimental spectra (Fig.3),  with some specific choice of the
wavelengths high-visibility polarization interference is observed
when rotating the polarization prism in channel 2. This is
well-known for the case where the selected wavelength is the central
one. Indeed, by selecting the wavelength of $702$ nm and rotating
the Glan prism in channel 2 we observed polarization interference
with a visibility of $89\%$. However, we see that high-visibility
polarization interference also takes place when the selected
wavelength is $695.5$ nm or $708.5$ nm. Both cases correspond to the
selection of the $|\Psi^-\rangle$ state.

To demonstrate polarization interference for the $|\Psi^-\rangle$
state, the wavelength transmitted by the monochromator was fixed at
$708.5$ nm and the coincidence counting rate was measured depending
on the orientation of polarizer 2, the other polarizer being
oriented at $\Theta_1=45^\circ$. The dependence, shown in Fig.4,
demonstrates a visibility of 98\%.

Finally, we have checked the invariance of the produced
$|\Psi^-\rangle$ state under polarization transformations, which
were performed by placing quarter- and half- wave plates (QWP and
HWP) at various orientations in front of the beamsplitter. For each
setting of the plates, the spectra similar to the ones shown in
Fig.3 were obtained. In particular, the dependencies obtained for
various HWP orientations are shown in Fig.5(a,b). As expected from
the theoretical prediction, the coincidence counting rate
corresponding to the wavelengths $695.5$ nm and $708.5$ nm (the
$|\Psi^-\rangle$ state)  at both positions of the Glan prisms,
($45^{\circ},45^{\circ}$) and ($45^{\circ},-45^{\circ}$), does not
change depending on the HWP orientation, i.e., depending on the
rotation of the two-photon polarization state before the
beamsplitter. The resulting visibility of polarization interference
exceeds 84\% for all settings of the HWP. At all other wavelengths
(including the central one, $702$ nm), the coincidence counting rate
clearly depends on the HWP orientation, and for some orientations,
the visibility of polarization interference becomes zero. Similar
behavior is observed for a QWP inserted before the beamsplitter.

The invariance of the $|\Psi^-\rangle$ state to polarization
transformations is also clearly seen when an optical fibre is used
as a spectral-selecting element. Indeed, the polarization drift,
unavoidably present in optical fibres, 'deteriorates' all the Bell
states except the singlet one $|\Psi^-\rangle$. To demonstrate this,
we have removed the monochromator and selected the singlet state by
means of a $1$ km fibre inserted before the beamsplitter.
Coincidence distributions obtained for the ($45^{\circ},45^{\circ}$)
and ($45^{\circ},-45^{\circ}$) settings of the Glan prisms are
presented in Fig.6. One can see that at time delays not
corresponding to the generation of the singlet state, the
dependencies have complicated shapes and do not correspond to the
behavior shown in Fig.3 (a result of polarization drift in the
fibre). At the same time, at points pertaining to the
$\left|\Psi^-\right>$ generation there is a coincidence minimum for
the ($45^{\circ},45^{\circ}$) settings and a maximum for the
($45^{\circ},-45^{\circ}$) settings of the prisms.

Using an optical fibre as a frequency-selecting element in the
analysis of the two-photon spectral amplitude is most convenient
whenever polarization drift can be eliminated or made inessential.
In particular, one can use a Faraday mirror (which acts as a time
reversal) and make the two-photon light travel twice through the
same fibre, thus compensating for the polarization drift. This
technique, suggested in~\cite{Faraday mirror}, was used for
two-photon spectral amplitude selection in~\cite{Brida}.

Another way to avoid the influence of polarization drift in the
fibre is to insert a single Glan prism before the fibre and no
prisms after the fibre. Then the role of both Glan prisms is played
by a single one, oriented at $45^{\circ}$; the payoff for
eliminating polarization drift is the impossibility to study the
coincidence spectrum in the $(45^{\circ},-45^{\circ})$
configuration. Using this technique, with a $1$ km fibre we
demonstrated the action of birefringent plates on the state produced
at various wavelength of SPDC bandwidth (Fig.7).

According to the considerations given in Section II, if a
birefringent plate introducing an e-o delay $\tau$ is inserted after
the crystal, the spectral dependencies (\ref{thetas}) become

\begin{eqnarray}
R_c(\Omega)=\hbox{sinc}^2(\Omega\tau_0)[\sin^2(\Theta_1+\Theta_2)\cos^2(\Omega(\tau_0+\tau))
\nonumber
\\
+\sin^2(\Theta_1-\Theta_2)\sin^2(\Omega(\tau_0+\tau))].
\label{thetasplates}
\end{eqnarray}

As mentioned above, for a complete compensation of the e-o delay,
$\tau=-\tau_0$, the state produced is $\left|\Psi^+\right>$ over the
whole bandwidth. Here we demonstrate the opposite effect, where
$\tau$ is of the same sign as $\tau_0$, and the fine structure of
the SPDC line becomes even finer. In particular, when $\tau=\tau_0$,
the modulation period in (\ref{thetasplates}) becomes twice as large
as in (\ref{thetas}), and $\left|\Psi^+\right>$ is produced at the
same pair of frequency offsets $\Omega=\pm\pi/2\tau_0$ where
$\left|\Psi^-\right>$ is produced in the absence of the birefringent
plate.

Indeed, in Fig.7 one can see that the modulation of the SPDC
lineshape observed in coincidences becomes more and more frequent as
we pass from no plates inserted (Fig.7a) to one quartz plate of
thickness $1$ mm (Fig.7b) and two quartz plates of thickness $1$ mm
(Fig.7c) inserted after the BBO crystal.

\subsection{III.2 'Two type-I' configuration.}

With the same experimental setup (and the frequency selection
performed by a monochromator), we have analyzed the frequency
structure of polarization-entangled states produced by two 1 mm
type-I BBO crystals cut for collinear frequency-degenerate
phasematching with the optic axes oriented in orthogonal planes. The
polarization of the pump was rotated to $45^\circ$ by a half-wave
plate, thus providing equal contributions from each crystal. The
phase between the contributions from both crystals at the central
wavelength $702$ nm was set to zero by tilting two quartz plates
with optic axes oriented vertically, placed into the two-photon
beam. As it follows from Eq. (\ref{Psi_Ifreq}), in this case the
lineshape of SPDC contains a continuous set of
polarization-entangled states from $\left|\Phi^+\right>$ at the
center to $\left|\Phi^-\right>$ at the slopes, with ${|HH \rangle +
e^{i \theta} | VV \rangle \over \sqrt 2} $ ($0<\theta < \pi)$
entangled states between them. When the orientations of the prisms
$\Theta_1,\Theta_2$ are varied, maximum visibility of polarization
interference is achieved for the Bell states
$\left|\Phi^\pm\right>$; the coincidence counting rate depends on
$\Theta_1,\Theta_2$ as $R_c\sim\cos^2(\Theta_1\mp\Theta_2)$.

Spectral distributions of coincidences for the
$(45^{\circ},45^{\circ})$ and $(45^{\circ},-45^{\circ})$
configurations of the Glan prisms are shown in Fig.8. The
theoretical dependence, similarly to (\ref{thetas}), can be written
as

\begin{eqnarray}
R_c(\Omega)=\hbox{sinc}^2
\left(\frac{d^2k_{o}}{d\Omega^2}\frac{\Omega^2L}{2}\right)\nonumber
\\
\times[\cos^2(\Theta_1-\Theta_2)\cos^2\left(\frac{d^2k_{e}}{d\Omega^2}\frac{\Omega^2L}{2}\right)
\nonumber
\\
+\cos^2(\Theta_1+\Theta_2)\sin^2\left(\frac{d^2k_{e}}{d\Omega^2}\frac{\Omega^2L}{2}\right)].
\label{thetasI}
\end{eqnarray}

We see that the Bell states $\left|\Phi^+\right>$ and
$\left|\Phi^-\right>$ are produced when the wavelengths selected are
the central one ($702$ nm) and the side ones ($658$ nm or $746$ nm),
respectively. Polarization interference for these states was
observed by selecting the corresponding wavelength, fixing one of
the Glan prisms at $\Theta_1=45^{\circ}$, and rotating the other one
(Fig.9).

Note that although it is commonly believed that the `two type I
crystals' scheme for polarization-entangled production requires
neither delay compensation nor filtering, from Fig.~8 one can see
that on the slopes of the spectrum, the state produced is different
from the one generated in the middle. Note that this occurs even
when perfectly collinear SPDC beam ($\theta=0$) is selected. Hence,
filtering the SPDC linewidth is also necessary in this case,
although the bandwidth of the filter can be rather large (from
Fig.~8, about $30$ nm).

\section{IV. Experiment: polarization-wavevector entanglement}

After analyzing frequency structure of the entanglement in type-II
or 'two type-I' configurations of SPDC, let us pass to considering
its angular structure. It was studied by means of the experimental
setup shown in Fig.10 where angular selection was performed in one
output port of the beamsplitter. For this purpose, we used a lens
with the focal length $500$ mm, and the detector apparatus placed in
its focal plane. The detector apparatus included an avalanche
photodiode, an interference filter with FWHM=$3$ nm, centered at the
degenerate wavelength $702$ nm, a lens with focal length $25$ mm,
and a pinhole with  $1$ mm diameter. Thus, the angular resolution in
the angle-selective arm of the setup (arm 1) was $2$ mrad. It is
important that, since the detection apparatus was placed into the
far-field zone, transverse walk-off effects did not contribute to
the results.

The other arm of the setup (arm 2) was designed  to collect all
angular width of SPDC spectrum corresponding to the degenerate
frequency. For this purpose, we removed the lens from the detection
apparatus of arm 2 and imaged the pump beam waist ($1.5$ mm) onto
the sensitive area of the avalanche photodiode ($200\mu\times
200\mu$) with a demagnification of $1:6$. This was obtained by using
an objective lens with a large numerical aperture and with the focal
length $F=100$ mm. The angular bandwidth of SPDC after the lens,
although increased $6$ times, was still within the acceptance angle
of the photodiode. In order to prevent a significant contribution of
accidental coincidences, an interference filter with FWHM=$1$ nm and
a vertical slit of diameter $2$ mm were placed before the detector.

\subsection{IV.1 Type-II configuration.}

In the case of type-II SPDC from a $0.5$ mm BBO crystal, we studied
the coincidence counting rate distribution using two geometries of
experiment. In the first geometry, the angle was scanned in the
plane of the optic axis. The corresponding dependence of the
coincidence counting rate on the angle selected by the pinhole in
arm 1 of the setup, for the $(45^{\circ},45^{\circ})$ and
$(45^{\circ},-45^{\circ})$ settings of the Glan prisms, is shown in
Fig.11. The dependencies are well fitted by the relation

\begin{eqnarray}
R_c(\theta)=\hbox{sinc}^2 \left(\frac{B\theta L}{2} \right)[\sin^2
\left(\Theta_1+\Theta_2 \right)\cos^2\left(\frac{B\theta L}{2}
\right) \nonumber
\\
+\sin^2(\Theta_1-\Theta_2)\sin^2 \left(\frac{B\theta L}{2}\right)],
\label{thetang}
\end{eqnarray}
which follows from Eq.(\ref{IIang}). Similarly to the case of the
frequency spectrum,  the state generated at the center of the
angular lineshape is $\left|\Psi^+\right>$, while at the angles
$\pm0.012$ rad the state $\left|\Psi^-\right>$ is produced. Indeed,
by selecting the angle in arm 1 to be $\theta=0$, fixing the Glan
prism in arm 1 at $45^{\circ}$ and rotating the other Glan prism we
obtained a polarization interference typical of
$\left|\Psi^+\right>$, with a visibility of $96.3\%$. At the same
time, the same measurement performed for the angle $\theta=0.01$ rad
selected, showed  a polarization-interference dependence typical of
$\left|\Psi^-\right>$, with the visibility $97\%$ (Fig.12).

In the other geometry of experiment, the angle of scattering was
scanned in the plane orthogonal to the plane of the optic axis. As
mentioned in Section II, in this case no 'fine structure' of the
angular lineshape should be observed since the wavevector mismatch
$\Delta_z$ is symmetric with respect to the interchange of the
wavevectors $k_o, k_{e}$. As a result, the produced state was
$\left|\Psi^+\right>$ for all angles of scattering $\theta$, and the
number of coincidences recorded for $(45^{\circ},45^{\circ})$
settings of the polarizers by far exceeded the one obtained for
$(45^{\circ},-45^{\circ})$ settings~(Fig.13).

\subsection{IV.2 'Two type-I' configuration.}

With a similar experimental setup (Fig.10) we performed the
measurements of the spatial spectra in the configuration with two
type I crystals with orthogonal optic axes. It is worth mentioning
that in this case it is critical that two production crystals are
mounted face to face. This arrangement allows to avoid the spatial
walk-off between the correlated beams produced from different
crystals which leads to their spatial distinguishability. Notice
that the problem of the walk-off is absent for the collinear
operation. Similar to the previous section we studied the
coincidences dependence on the angle selected by the pinhole at a
fixed orientations of the polarizers P1 and P2. The quartz plates
installed in the pump beam were aligned in a way that the entangled
$\left|\Phi^+\right>$ state was generated in the collinear
configuration. If the pinhole filters a small angle $\theta$ within
the line of SPDC then the generated state is given by (10). In this
case the dependence of the coincidences counting rate is given by
(similar to (14)):

\begin{eqnarray}
R_c(\theta)=\hbox{sinc}^2
\left(\frac{k_{o}\theta^2L}{2}\right)[\sin^2(\Theta_1+\Theta_2)\cos^2\left(\frac{k_{e}\theta^2L}{2}\right)
\nonumber
\\
+\sin^2\left(\Theta_1-\Theta_2\right)\sin^2\left(\frac{k_{e}\theta^2L}{2}\right)].
\label{thetang}
\end{eqnarray}

The results of the experiment for the orientations of the polarizers
at $(45^{\circ},45^{\circ})$ and $(45^{\circ},-45^{\circ})$ are
presented in Fig.14 . As one expects from the theoretical
considerations (see section II) whilst an entangled
$\left|\Phi^+\right>$ state is generated within the wide angular
bandwidth near the collinear configuration, at side-bands of the
spectra the state is transformed into the orthogonal
$\left|\Phi^-\right>$ state.

\section{V. Conclusion}

Finally, let us drive some conclusions from the obtained theoretical
and experimental results.

Having analyzed two possible schemes for producing
polarization-entangled states via cw-pumped SPDC in a crystal of
finite length $L$ (usual type-II phasematching and the 'two type-I'
scheme), we see that within the natural bandwidth, there is never a
single Bell state produced for all mismatch values. At angles and
frequencies corresponding to exact phase matching, the Bell state
$\left|\Psi^+\right>$ is produced for type-II SPDC and the Bell
state $\left|\Phi^+\right>$, for the 'two type-I' scheme. However,
at the values of the phase mismatch $\pm\pi/L$, the state produced
in these two schemes is $\left|\Psi^-\right>$ and
$\left|\Phi^-\right>$, respectively. At intermediate mismatches,
other maximally entangled states are produced. Since the mismatch
variation can be performed through the variation of either frequency
or angle of scattering, or both, there is a 'fine structure' of
maximally entangled states generation in the two-dimensional
frequency-angular spectrum of SPDC.

It is well-known that for type-II SPDC, one can  produce the state
$\left|\Psi^+\right>$ within the whole frequency bandwidth by
compensating  the delay between the two orthogonally polarized
photons of a pair born in the same crystal. This is usually done by
placing after the crystal a birefingent plate of appropriate
thickness. However, this compensation, although increasing the
efficiency of Bell-states generation, still leaves the fine
structure in the angular lineshape. To avoid the generation of other
Bell states, in all existing experiments one usually selects a
sufficiently small angular width by means of a pinhole. From our
consideration, it immediately follows that the fine structure of
Bell-states generation within the angular SPDC linewidth can be
compensated similarly, by placing an anisotropic crystal into the
SPDC beam. This would lead to a much more efficient Bell-states
generation compared to the case of selecting a narrow angular
spectrum.

The case of non-compensated fine structure of Bell states production
within the natural SPDC linewidth is also very interesting. An
important point that has been evidenced is that the frequency
structure can be made to vary more rapidly if the birefringent plate
after the crystal increases the e-o delay, instead of reducing it.
The advantage of having such a structure is that one can produce
different collinear Bell states at non-degenerate wavelengths. They
can find application in  realizing, for instance, quantum
information protocols involving the singlet state and one of the
triplet states, e.g.~\cite{Dragan}.

The fact that the frequency structure is rather narrow (for a
type-II crystal of thickness $0.5$ mm, it is on the order of ten
nanometers) is not a drawback. Indeed, if the Bell states produced
are to be applied for quantum communication, which is usually
performed through optical fibres, large wavelength separation should
be avoided because of the polarization-mode dispersion. At the same
time, we have shown that selection of the wavelength can be in this
case performed by means of the fibre itself, through the time
selection of the two-photon amplitude.

The possibility to select a Bell state by means of appropriate
angular filtering may be of considerable importance since the
angular filtering can be principally lossless. This can be used in
applications where losses are crucial such as, for instance,
squeezed states generation.

The observed fine structure of SPDC lineshape can be also applied in
quantum information for the preparation of complex entangled states,
for instance, states with double entanglement in frequency,
wavevector, and polarization.

Finally, it is worth mentioning that, although here we only
considered collinear frequency-degenerate SPDC, the same effects are
also present in the non-collinear or frequency non-degenerate cases.

\textbf{Acknowledgements:} We would like to acknowledge Marco Terzi,
Ivo DeGiovanni, Luca Giacone and Valentina Caricato for fruitful
discussions and help with the experiment. This work was supported in
part by the joint grant RFBR-Piedmont 07-02-91581-ASP, by italian
minister of research (PRIN 2005023443-002) and by Regione Piemonte
(E14).

\begin{figure}
\includegraphics[height=5cm]{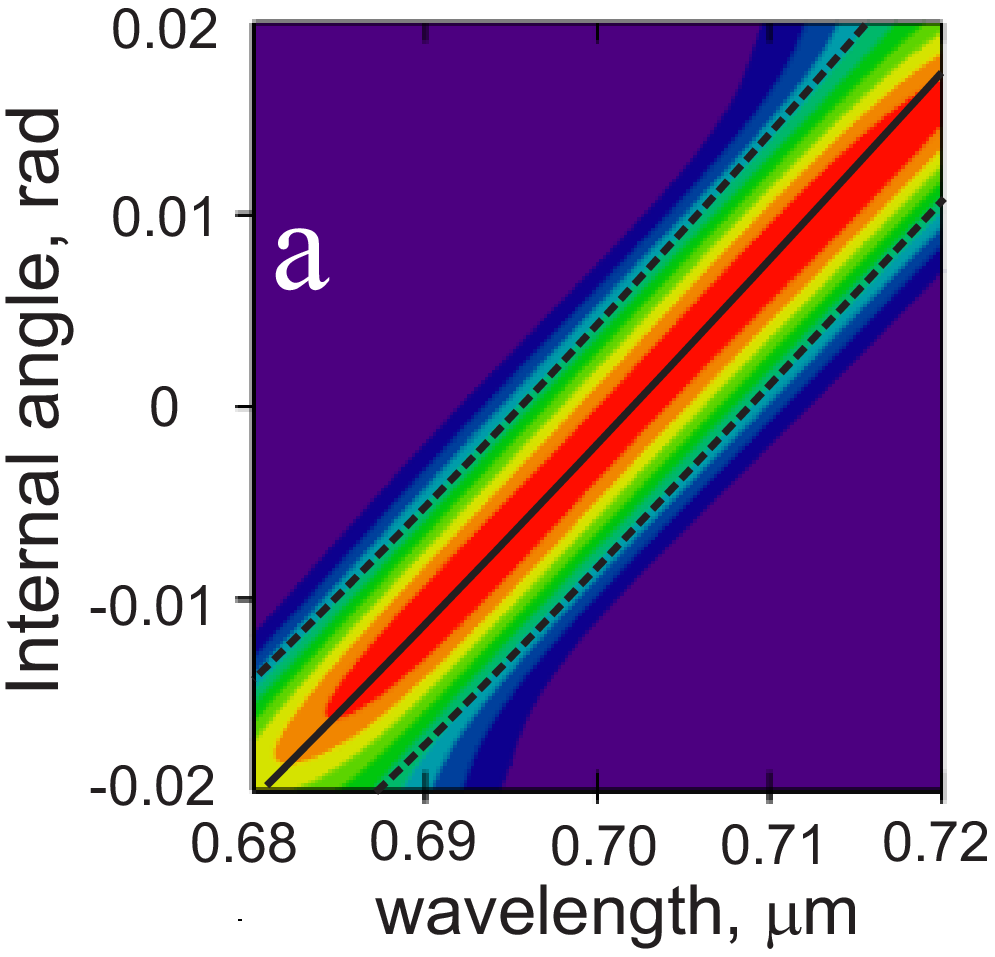}
\includegraphics[height=5cm]{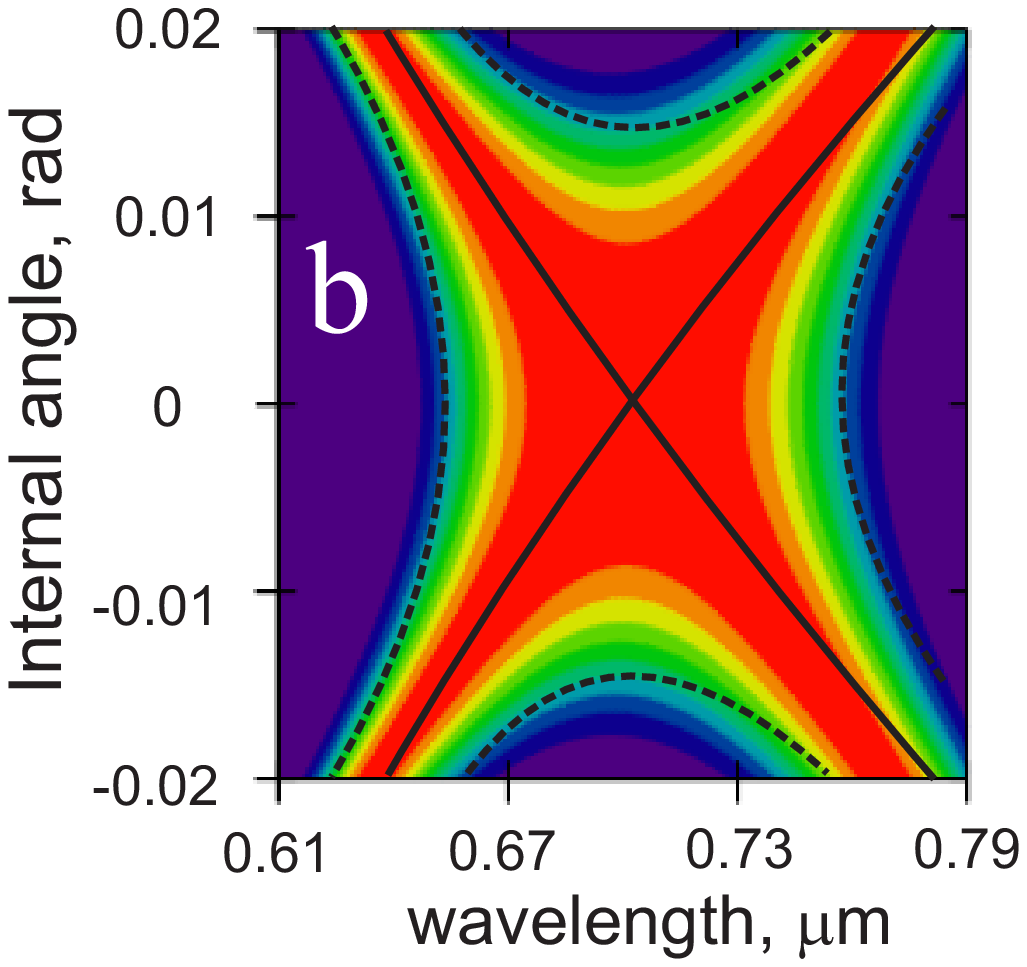}
\caption{Frequency-angular distributions of SPDC calculated for the
case of (a) type-II collinear frequency-degenerate phase matching in
a 0.5 mm BBO crystal and (b) two 1 mm type-I BBO crystals cut for
collinear frequency-degenerate phase matching, placed one after
another, with optic axes in orthogonal planes, and pumped by the
same laser beam (supposed to have a wavelength of 351 nm). Solid
lines in (a) show the domains where $\left|\Psi^+\right>$ is
generated, dashed lines, where $\left|\Psi^-\right>$ is generated.
Solid lines in (b) show the domains where $\left|\Phi^+\right>$ is
generated, dashed lines, where $\left|\Phi^-\right>$ is generated.}
\end{figure}

\begin{figure}
\includegraphics[height=4cm]{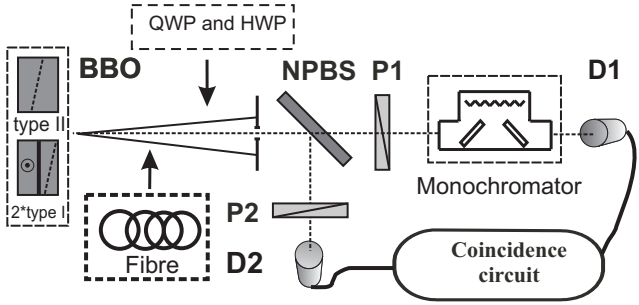} \caption{Scheme of the experimental setup.
A type-II BBO crystal or a combination of two type-I crystals cut
for collinear frequency-degenerate phasematching is pumped by cw
$Ar^+$ laser at 351 nm; NPBS, a 50/50 nonpolarizing BS; P1 and P2,
Glan prisms; D1, D2, single-photon counting modules. A
frequency-selecting element is placed in one arm; in the other arm,
all frequencies are collected. Retardation plates, QWP and HWP, are
used to study the invariance of the Bell state $|\Psi^-\rangle$
under polarization transformations. In some measurements, the
monochromator is replaced by a $1$ km single-mode fiber inserted
after the crystal and frequency selection is replaced by time
selection.}
\end{figure}

\begin{figure}
\includegraphics[height=5cm]{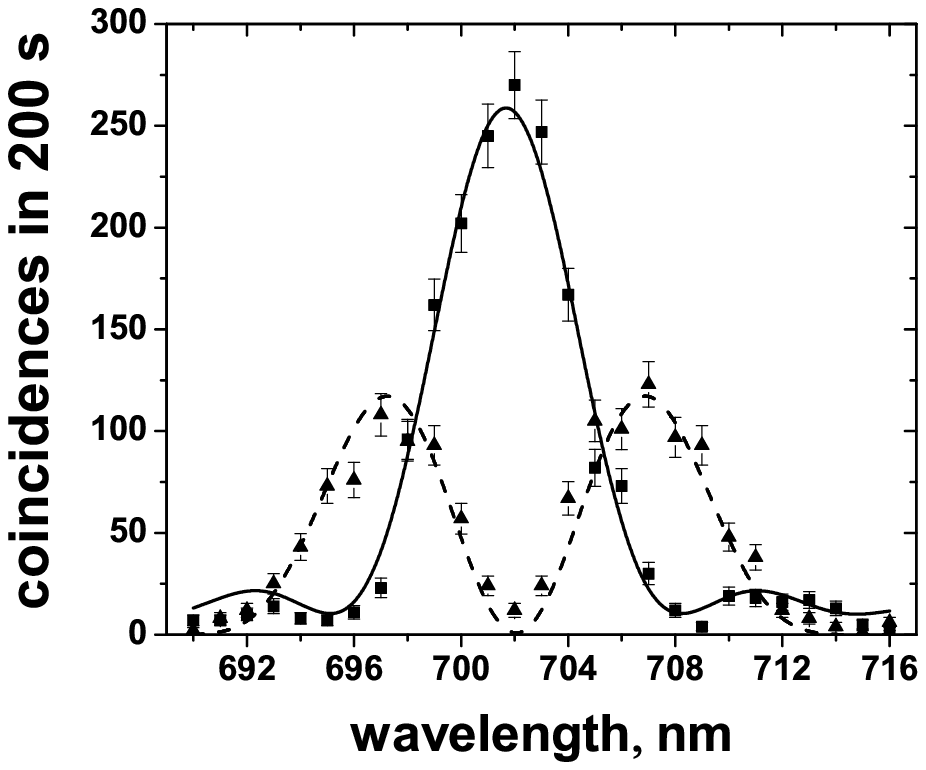}
\caption{Experimental dependence of the coincidence counting rate
for type-II SPDC on the wavelength selected by the monochromator for
two cases: $\Theta_1=\Theta_2=45^\circ$ (squares, solid line) and
$\Theta_1=45^\circ, \Theta_2=-45^\circ$ (triangles, dashed line).
Lines represent a fit with Eq.(\ref{thetas}).}
\end{figure}

\begin{figure}
\includegraphics[height=5cm]{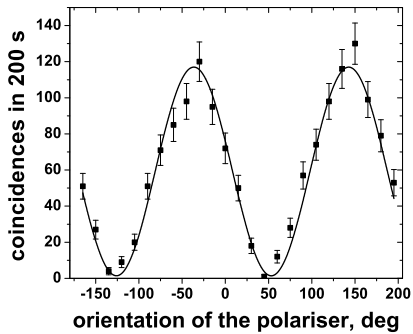}
\caption{Polarization interference fringes for the singlet Bell
state $|\Psi^-\rangle$ generated in type-II SPDC (the selected
wavelength is $\lambda=708.5$nm).}
\end{figure}

\begin{figure}
\includegraphics[height=5cm]{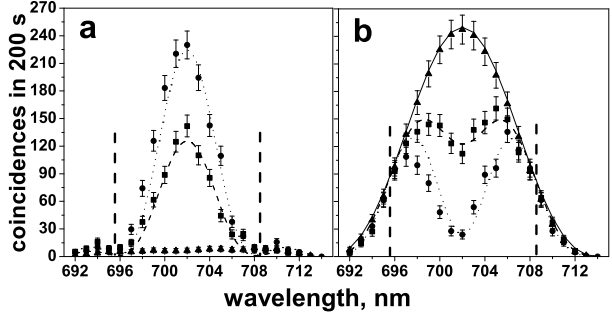}
\caption{Experimental dependence of the coincidence counting rate on
the wavelength selected by the monochromator for the following
orientations of the HWP placed after the crystal: $7^\circ$(circles,
dotted line); $17^\circ$ (squares, dashed line ); $22,5^\circ$
(triangles, solid line). Orientations of the Glan prisms are
$45^\circ,45^\circ$ (a) and $45^\circ,-45^\circ$ (b).
Dashed vertical bars show the wavelength where $|\Psi^-\rangle$ is
generated. Lines represent the theoretical fit.}
\end{figure}

\begin{figure}
\includegraphics[height=5cm]{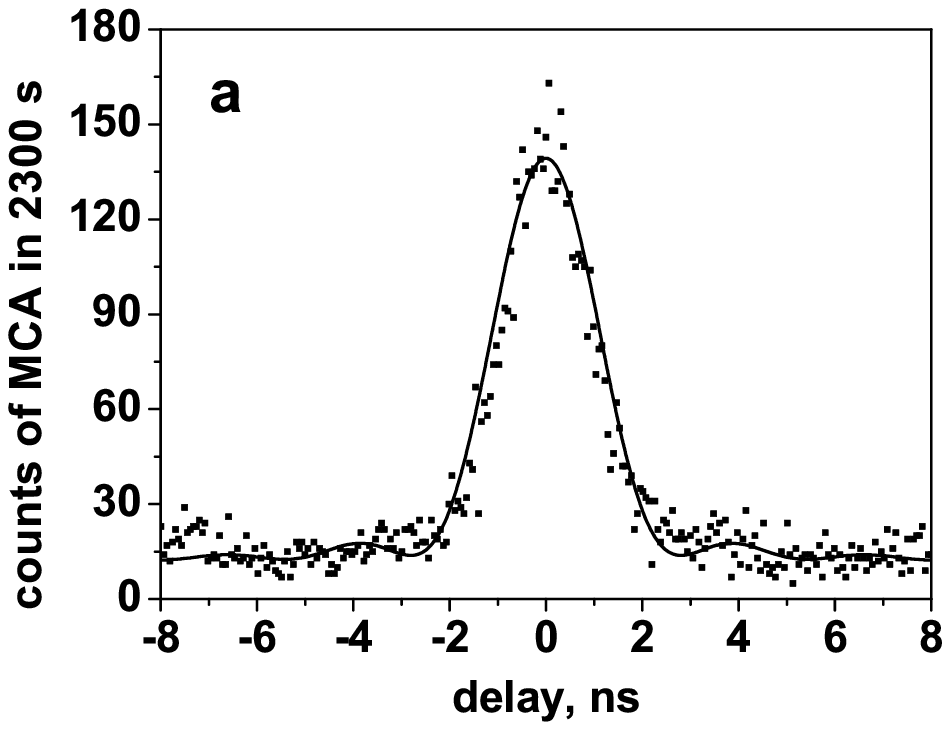}
\includegraphics[height=5cm]{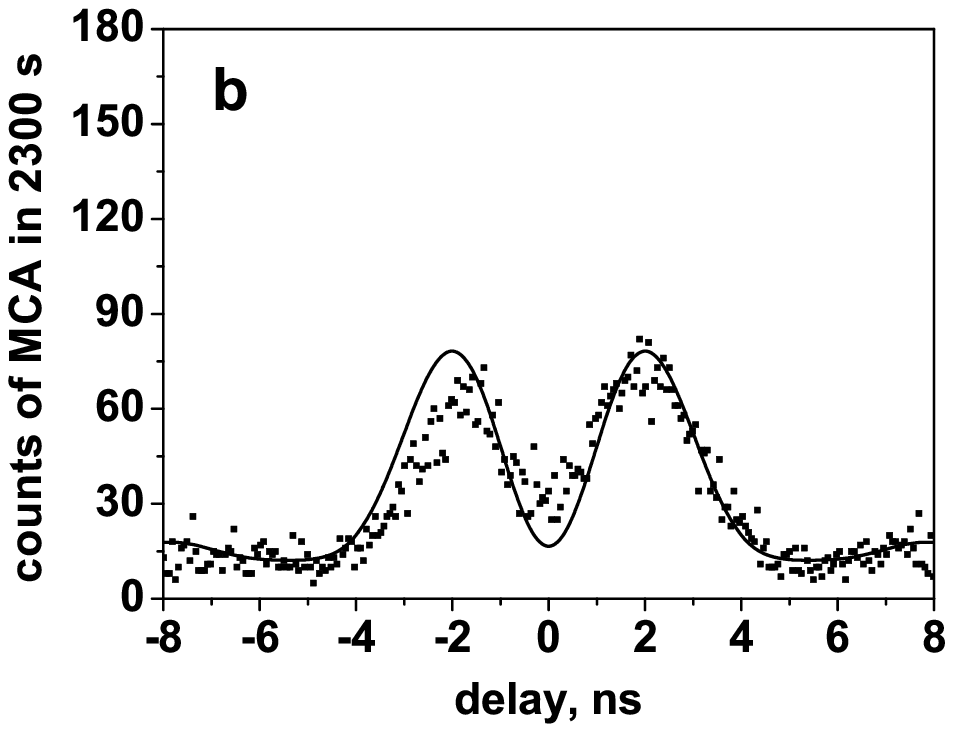}
\caption{Experimental dependence of the coincidence counting rate on
the delay between the arrivals of two photons, with the
monochromator replaced by a 1 km optical fibre and the Glan prisms
set at positions (a) ($45^{\circ},45^{\circ}$) and (b) ($45^{\circ},
-45^{\circ}$).}
\end{figure}

\begin{figure}
\includegraphics[height=5cm]{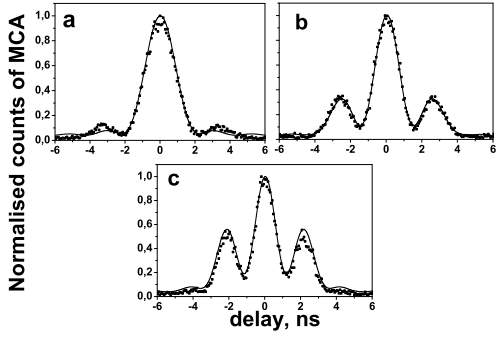}
\caption{The measured shape of the second-order correlation function
for the case of (a) no quartz plates, (b) one quartz plate of
thickness 1 mm, and (c) two quartz plates of thickness 1 mm placed
at the output of the crystal with the optic axes parallel to the
plane of the BBO optic axis, and a single polarizer set at
$\Theta=45^{\circ}$ before the fibre. The curves show the
theoretical dependencies (\ref{thetasplates}) with an account for
the detector jitter time.}
\end{figure}

\begin{figure}
\includegraphics[height=5cm]{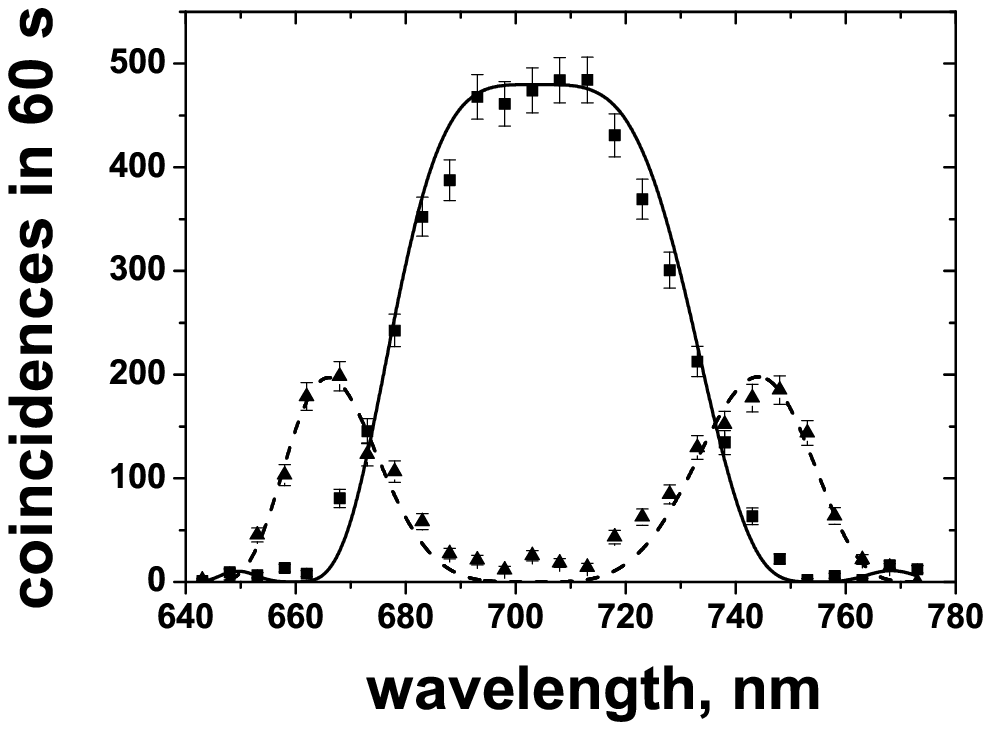}
\caption{Experimental dependence of the coincidence counting rate
for the 'two-type I' configuration on the wavelength selected by the
monochromator for two cases: $\Theta_1=\Theta_2=45^\circ$ (squares,
solid line) and $\Theta_1=45^\circ, \Theta_2=-45^\circ$ (triangles,
dashed line). Lines represent a fit with Eq.(\ref{thetasI}).}
\end{figure}

\begin{figure}
\includegraphics[height=5cm]{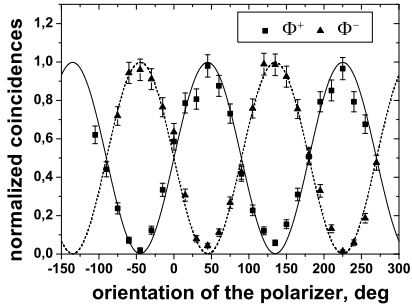}
\caption{Polarization interference fringes for the Bell states
$|\Phi^+\rangle$ (the selected wavelength is $\lambda=702$nm) and
$|\Phi^-\rangle$ (the selected wavelength is $\lambda=658$nm)
generated from the 'two type-I' configuration.}
\end{figure}

\begin{figure}
\includegraphics[height=4cm]{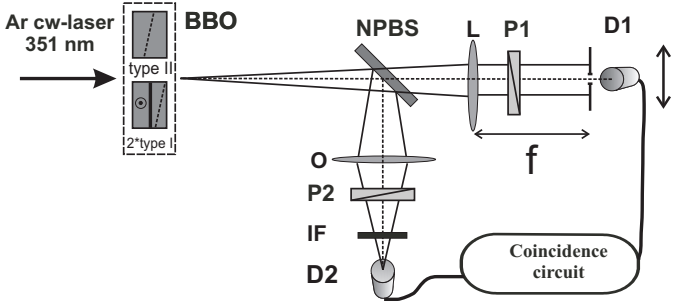}
\caption{Scheme of the experimental setup. The crystals arrangement
is the same as in fig.2 ; NPBS, a 50/50 nonpolarizing BS; P1 and P2,
Glan prisms; D1, D2, single-photon counting modules. Angular
selection is performed in one arm by a lens L with focal length
$500$ mm and a pinhole with diameter $1$ mm, placed in a focal plane
of L and scanned in the vertical plane together with the detector
D1. In the other arm, all angles are collected by means of an
objective O with focal length $100$ mm. Narrow-band interference
filter IF is used to avoid contribution of accidental coincidences.}
\end{figure}

\begin{figure}
\includegraphics[height=5cm]{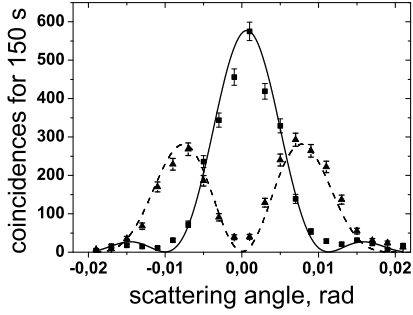}
\caption{Experimental dependence of the coincidence counting rate
for the 'type II' configuration on the angle selected by the pinhole
in arm 1 of the setup for two cases: $\Theta_1=\Theta_2=45^\circ$
(squares, solid line) and $\Theta_1=45^\circ, \Theta_2=-45^\circ$
(triangles, dashed line). Angular scanning was performed in the
plane of the optic axis. Lines represent a fit with
Eq.(\ref{thetang}).}
\end{figure}

\begin{figure}
\includegraphics[height=5cm]{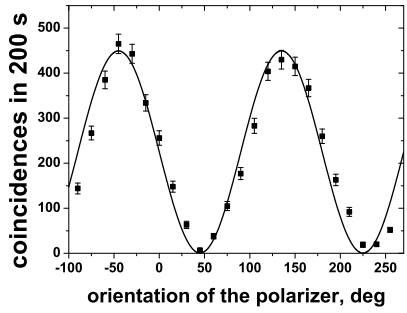}
\caption{Polarization interference fringes for the singlet Bell
state $|\Psi^-\rangle$ generated in type-II SPDC (the selected angle
is $\theta=-0.012$ rad).}
\end{figure}

\begin{figure}
\includegraphics[height=5cm]{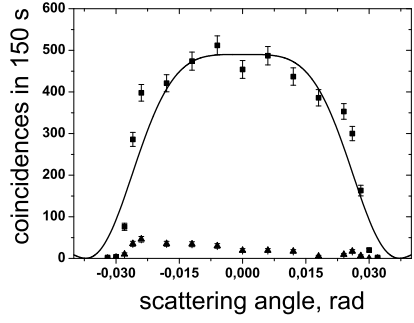}
\caption{Experimental dependence of the coincidence counting rate
for the 'type II' configuration on the angle selected by the pinhole
in arm 1 of the setup for two cases: $\Theta_1=\Theta_2=45^\circ$
(squares, solid line) and $\Theta_1=45^\circ, \Theta_2=-45^\circ$
(triangles, dashed line). Angular scanning was performed in the
plane orthogonal to the plane of the optic axis.}
\end{figure}

\begin{figure}
\includegraphics[height=5cm]{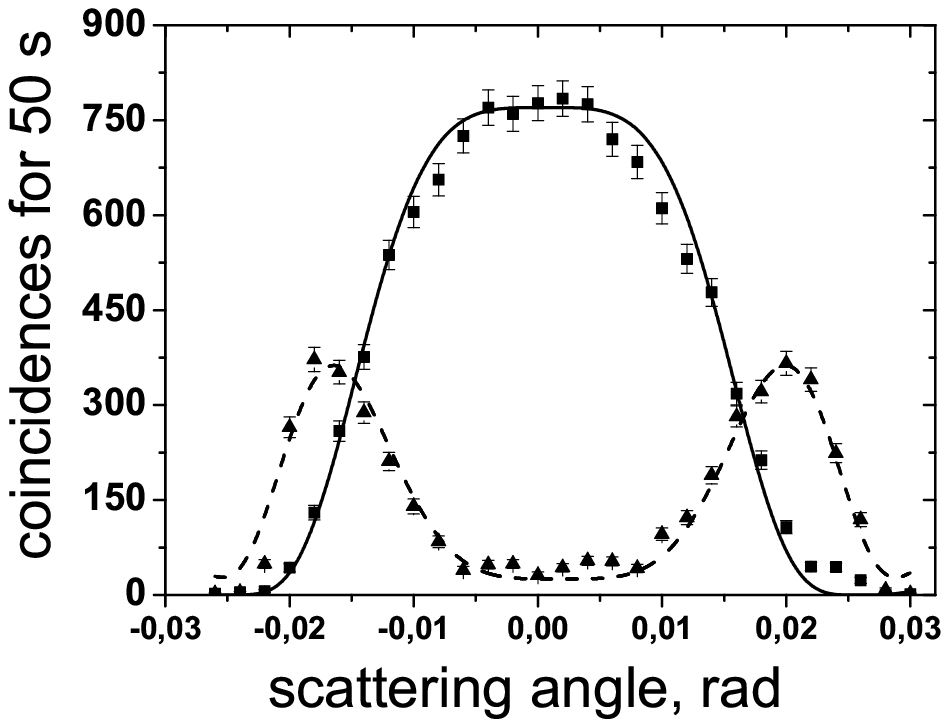}
\caption{Experimental dependence of the coincidence counting rate
for the 'two-type I' configuration on the angle selected by the
pinhole for two cases: $\Theta_1=\Theta_2=45^\circ$ (squares, solid
line) and $\Theta_1=45^\circ, \Theta_2=-45^\circ$ (triangles, dashed
line).}
\end{figure}


\begin{references}

\bibitem{NC} M.A. Nielsen and I.L. Chuang, "Quantum Computation and
Quantum Information", Cambridge Univ. press (Cambridge, 2000). D.
Bouwmeester et al., "The physics of Quantum Information", Springer
Verlag (Berlin, 2000).

\bibitem{prep} M.Genovese,  \emph{Phys. Rep}. \textbf{413} 319 (2005).

\bibitem{las} G. Brida, M.Genovese, M. Gramegna, Laser Physics Letters 3
(2006) 115 and Refs. therein.

\bibitem{qi} L.A. Lugiato, A. Gatti and E. Brambilla, J. Opt. B.
Quant. Sem. Opt. 4 (2002) S176.

\bibitem{au} A. Garuccio, in "Fundamental Problems in Quantum Theory", Ed. D. Greenberger ,
(New York Academy of Sciences, 1995).

\bibitem{Kwiat_rings} P.G.Kwiat,
K.Mattle, H.Weinfurter, A.Zeilinger, A.V.Sergienko, Y.H.Shih. Phys.
Rev. Lett. \textbf{75}, 4337 (1995).

\bibitem{Branning} D.Branning, W.~P.~Grice, R.~Erdmann, and
I.~A.~Walmsley, Phys. Rev. Lett. \textbf{83}, 955 (1999);
\bibitem{Kim} Y.-H.~Kim, S.~P.~Kulik, M.~V.~Chekhova, W.~P.~Grice,
and Y.~H.~Shih,Phys. Rev. A \textbf{67}, 010301(R) (2003).


\bibitem{Kwiat} P.G.Kwiat,
E.Waks, A.G.White, I.Appelbaum, and P.G.Eberhard, Phys. Rev. A
\textbf{60},R773 (1999); G. Brida et al., Phys. Lett. A\textbf{268},
12 (2000); A\textbf{299},  121 (2002).

\bibitem{Sergei+Yoon-Ho} Y.H.Kim, S.P.Kulik, and Y.H.Shih. Phys. Rev. A \textbf{63},
060301(R) (2001).

\bibitem{Yura} A.V.Burlakov,
S.P.Kulik, G.O.Rytikov, and M.V.Chekhova, JETP \textbf{95}, 639-644
(2002).

\bibitem{Shih} M. H. Rubin, D. N. Klyshko, Y. H. Shih, and A. V. Sergienko,
Phys. Rev. A 50, 5122 (1994).

\bibitem{FWM} L.J.Wang, C.K.Hong, and S.R.Friberg, J. Opt. B: Quantum and Semiclass. Opt. \textbf{3},
346 (2001); J. E. Sharping, M. Fiorentino, and P. Kumar, Opt. Lett.
26, 367–369 (2001); H.Takesue and K.Inoue, Phys. Rev. A 70,
031802(R) (2004); J.G.Rarity \emph{et al.}, Optics Express,
\textbf{13}, No 2, 534 (2005).

\bibitem{DNK} D.N.Klyshko, \emph{Photons and Nonlinear Optics} (Gordon and Breach, New York,
1988).

\bibitem{BigDNK} A. V. Burlakov, M. V. Chekhova, D. N. Klyshko, S. P. Kulik,
A. N. Penin, Y. H. Shih, and D. V. Strekalov, Phys. Rev. A 56,
3214–3225 (1997).

\bibitem{Brida} G.~Brida, M.~V.~Chekhova, M.~Genovese, and L.~A.~Krivitsky,
Phys. Rev. Lett. \textbf{96}, 143601 (2006); Phys. Rev. A
\textbf{75}, 015801 (2007); Journ. of Phys. A, in press.

\bibitem{Faraday mirror}Martinelli M. 1992 \emph{Journ. Mod. Opt.} \textbf{39}
451; Martinelli M. 1989 \emph{ Opt. Comm.} \textbf{72} 341; Breguet
J. and Gisin N. 1995 \emph{Opt. Lett.} \textbf{20}1447.


\bibitem{Dragan} K.Banaszek, A.Dragan, W.Wasilewski, and Czeslaw Radzewicz.
Phys. Rev. Lett. \textbf{92},
257901 (2004).


\end{references}
\end{document}